\documentclass[aip,amsmath,amssymb,reprint]{revtex4-1}

\usepackage{graphicx}
\usepackage{dcolumn}
\usepackage{bm}
\usepackage{textcomp}
\usepackage[utf8]{inputenc}
\usepackage[T1]{fontenc}
\usepackage{mathptmx}
\usepackage{appendix}
\usepackage{color}
\usepackage{textcomp}
\usepackage{subfigure}
\usepackage[colorlinks,citecolor=blue,linkcolor=red,hyperindex,CJKbookmarks]{hyperref}

\begin{document}
\title{Enantio-discrimination via light deflection effect}

\author{Yu-Yuan Chen}
\affiliation{Beijing Computational Science Research Center, Beijing 100193, China}

\author{Chong Ye}
\affiliation{Beijing Computational Science Research Center, Beijing 100193, China}

\author{Quansheng Zhang}
\affiliation{Beijing Computational Science Research Center, Beijing 100193, China}

\author{Yong Li}
\email{liyong@csrc.ac.cn}
\affiliation{Beijing Computational Science Research Center, Beijing 100193, China}
\affiliation{Synergetic Innovation Center for Quantum Effects and Applications, Hunan Normal University, Changsha 410081, China}
\date{\today}

\begin{abstract}
We propose a theoretical method for enantio-discrimination based on the light deflection effect in four-level models of chiral molecules. This four-level model consists of a cyclic three-level subsystem coupled by three strong driving fields and an auxiliary level connected to the cyclic three-level subsystem by a weak probe field. It is shown that the induced refractive index for the weak probe field is chirality-dependent. Thus it will lead to chirality-dependent light deflection when the intensities of two of the three strong driving fields are spatially inhomogeneous. As a result, the deflection angle of the weak probe light can be utilized to detect the chirality of pure enantiomers and enantiomeric excess of chiral mixture. Therefore, our method may act as a tool for enantio-discrimination.
\end{abstract}

\pacs{11.30.Rd, 33.80.-b, 42.50.-p}
\maketitle

\section{Introduction}
In nature, one of the most important manifestations of symmetry breaking is the separated existence of two molecular structural forms known as enantiomers (e.g., $R$ and $S$ enantiomers) which are mirror images of each other~\cite{MolecularStructure}. Molecular chirality has received considerable interest due to its fundamental role in the activity of various biological molecules and chemical reactions~\cite{EnantioEffect-Science1991,EnantioEffect-Nature1997}. Despite this importance, detecting the chirality of pure enantiomer (containing only $R$ or $S$ enantiomers) and enantiomeric excess of chiral mixture (containing both $R$ and $S$ enantiomers) remains an important and challenging task~\cite{Challenge-CD,Challenge-VCD,
Challenge-ROA,challenge-Nature1998,challenge-Nature2000,Challenge-Science2002,Challenge-JSepSci2007}.

The optical rotation, which refers to the rotation of the polarization plane of a linearly polarized light in a medium of chiral molecules, is one of the conventional spectroscopic methods for enantio-discrimination (detecting the chirality of pure enantiomers and enantiomeric excess of chiral mixture)~\cite{OR-Science1998,OR-Chirality2003,2013WeakMeasOR,2016WeakMeasOR+CCD,2018WeakMeasOR}. The optical rotation arises from the fact that the opposite enantiomers (e.g., $R$ or $S$ enantiomers) can lead to different refractive index for left- and right-circularly polarized lights~\cite{Barron-MolecularScatter}. Inspired by this, the reflection, refraction, and diffraction effects of left- and right-circularly polarized lights have been proposed to detect the chirality of pure enantiomers and enantiomeric excess of chiral mixture in experiments~\cite{ChiralReflectRefract,ChiralDiffract,Enhanced-ChiralRefract}. Nevertheless, the chiral effects in most of these methods are based on the interference between the electric- and magnetic-dipole transitions and thus are weak since the magnetic-dipole transition moment is usually weak.

Recently, the cyclic three-level ($\Delta$-type) model~\cite{LiuYuXi-SQClooplevel,YeChong-RealLoop} of chiral molecules based on electric-dipole transitions has received much interest and has been widely used in enantio-separation~\cite{LY-2007SeparationLoop,LY-2008SeparationLoop,Shapiro-2010SeparationLoop,JiaWZ-2010SeparationLoop,
Vitanov-2019SeparationLoop,WuJL-2019SeparationLoop} and enantio-discrimination \cite{JiaWZ-2011DiscriminationLoop,
Hirota-2012DiscriminationLoop,Doyle-2013DiscriminationLoop,Doyle-2013ChiralTWMLoop,Doyle-2014ChiralTWMLoop,
Lehmann-2015DiscriminationLoop,Schnell-2014DiscriminationLoop,Schnell-2015DiscriminationLoop,
Lehmann-2017DiscriminationLoop}. In such a model, the product of three Rabi frequencies in the cyclic three-level model can change sign with enantiomer, thus it will lead to chirality-dependent dynamic processes~\cite{LY-2008SeparationLoop,Vitanov-2019SeparationLoop,Hirota-2012DiscriminationLoop,
Doyle-2013DiscriminationLoop,Doyle-2013ChiralTWMLoop,Doyle-2014ChiralTWMLoop,
Lehmann-2015DiscriminationLoop,Schnell-2014DiscriminationLoop,Schnell-2015DiscriminationLoop,
Lehmann-2017DiscriminationLoop,Doyle-2017decoherenceValue} and steady-state optical responses~\cite{JiaWZ-2011DiscriminationLoop}.

In this paper, we propose a theoretical method for enantio-discrimination based on the light deflection effect in four-level models of chiral molecules. This four-level model consists of a cyclic three-level subsystem coupled by three strong driving fields and an auxiliary level which is connected to the cyclic three-level subsystem by a weak probe field. Here, two of the three strong driving fields are spatially inhomogeneous in order to produce spatially inhomogeneous refractive index for the weak probe light. Since the product of three Rabi frequencies corresponding to the cyclic three-level subsystem can change sign with enantiomer, the resultant refractive index leads to chirality-dependent propagation trajectory of the weak probe light which propagates in a medium of pure enantiomers. Therefore, the chirality of pure enantiomers can be detected by monitoring the deflection angle of the weak probe light when it exits from the medium. Further, we demonstrate that such a chirality-dependent deflection angle can also be utilized to detect the enantiomeric excess of chiral mixture. Moreover, the tunability of the amplitude of the deflection angle and robustness of the measurement precision against the Rabi frequency corresponding to the homogeneous driving field and enantiomeric excess are also investigated.

Here, we remark that the principle of our method is to measure the deflection angle of a probe light when it exits from the medium of enantiomers, which is very different from that of most optical rotation based methods~\cite{OR-Science1998,OR-Chirality2003,Barron-MolecularScatter,ChiralReflectRefract,ChiralDiffract,Enhanced-ChiralRefract,2013WeakMeasOR,2016WeakMeasOR+CCD,2018WeakMeasOR}. Moreover, our method involves only the electric-dipole transitions and thus can produce desirable chirality-dependent deflection angle. Hence, our method may have wide applications in enantio-discrimination.

This paper is organized as follows: In Sec.~\ref{detectchirality}, we derive the deflection angle of probe light which propagates in a medium of pure enantiomers based on the four-level model and illustrate how to detect the chirality of pure $R$ and $S$ enantiomers via such a deflection angle. In Sec.~\ref{detectEE}, we investigate the light deflection in a medium of chiral mixture and show the results of detecting the enantiomeric excess of chiral mixture. The discussion about our investigation is presented in Sec.~\ref{discussion}. Finally, the conclusion is given in Sec.~\ref{summary}.

\section{DETECTING THE CHIRALITY OF PURE ENANTIOMERS}\label{detectchirality}
The principle of our method depends on the mechanism of light deflection which is an important technology in modern optics~\cite{Gottlieb-DeflectTech,Holzner-DeflectTechExp,Moseley-DeflectTech,Enhance-deflect}. Recently, light deflection in homogeneous medium subject to inhomogeneous external fields has attracted much attention~\cite{ZhouLan-2007HomoDeflection,Kumar-2016HomoDeflection,ZhouLan-2008HomoDeflection}. It has potential applications in many fields such as steering, splitting, focusing, and cloaking of optical beam~\cite{Zubairy-2006DeflectionAppl,ZhouLan-2009DeflectionAppl,Scully-2010DeflectionAppl,
ZhuChengJie-2013DeflectionAppl,Verma-2015DeflectionAppl} due to the achievement in significant deflection angle. In order to investigate the light deflection phenomenon in a gaseous medium of orientated pure enantiomers, we begin with the master equations describing the dynamical evolution of the four-level model of $R$ or $S$ enantiomers, and then derive the chirality-dependent deflection trajectory in a medium of pure $R$ ($S$) enantiomers based on the steady-state optical response which can be determined by solving the master equations.

\subsection{Four-level model of chiral molecules}
\begin{figure}[tbp]
\centering
\includegraphics[width=8.5cm]{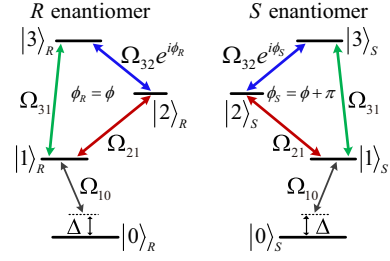}
\caption{(Color online) The schematic diagram of the four-level model for pure $R$ and $S$ enantiomers coupled by three strong driving fields and a weak probe field. The initially populated states are $\left|0\right\rangle_{L}$ and $\left|0\right\rangle_{R}$, respectively.}
\label{fig1}
\end{figure}

Each of the two opposite enantiomers can be modeled simultaneously as the four-level chiral-molecule model consisting of a cyclic three-level subsystem and an auxiliary level as shown in Fig.~\ref{fig1}. Here, the indices $R$ and $S$ are introduced to represent respectively the $R$ and $S$ enantiomers which are mirror symmetry of each other. $\left|k\right\rangle _{R}$ and $\left|k\right\rangle _{S}$ ($k=0,1,2,3$) are respectively the desired $k$-th eigen-states of $R$ and $S$ enantiomers. Here, we neglect the parity violating energy differences due to the fundamental weak force~\cite{PVED-PRL2000,PVED-PRA2001,
PVED-JCP2001}, thus $\left|k\right\rangle _{R}$ and $\left|k\right\rangle _{S}$ are with the same eigen-energy $\hbar\omega_{k}$. Moreover, we also neglect any tunneling between enantiomers. Three strong driving fields with amplitudes $E_{21}$, $E_{31}$, and $E_{32}$ and frequencies $\nu_{21}$, $\nu_{31}$, and $\nu_{32}$, are applied to resonantly couple respectively the transitions $\left|2\right\rangle_{Q} $$\leftrightarrow$$\left|1\right\rangle_{Q}$, $\left|3\right\rangle_{Q} $$\leftrightarrow$$\left|1\right\rangle_{Q}$, and $\left|3\right\rangle_{Q} $$\leftrightarrow$$\left|2\right\rangle_{Q}$ ($Q=R,S$) via the electric-dipole couplings. Thus it forms the cyclic three-level subsystem among $\left|1\right\rangle_{Q}$, $\left|2\right\rangle_{Q}$, and $\left|3\right\rangle_{Q}$. Meanwhile, a weak probe field $E_{10}$ with frequency $\nu_{10}$ couples the ground state $\left|0\right\rangle_{Q}$ to the cyclic three-level subsystem via the transition $\left|1\right\rangle_{Q}$$\leftrightarrow$$\left|0\right\rangle_{Q}$.

Under the dipole approximation and rotating-wave approximation, the Hamiltonian in the interaction picture with respect to $H_{0}^{Q}=\hbar[(\omega_{0}+\Delta)\left|0\right\rangle_{Q Q}\hspace{-0.2em}\left\langle 0\right|+\omega_{1}\left|1\right\rangle_{Q Q}\hspace{-0.2em}\left\langle 1\right|+\omega_{2}\left|2\right\rangle_{Q Q}\hspace{-0.2em}\left\langle 2\right|+\omega_{3}\left|3\right\rangle_{Q Q}\hspace{-0.2em}\left\langle 3\right|]$ is given as
\begin{align}
H_{I}^{Q} & =-\hbar\Delta\left|0\right\rangle_{Q Q}\hspace{-0.2em}\left\langle 0\right|-\hbar[\Omega_{10}\left|1\right\rangle_{Q Q}\hspace{-0.2em}\left\langle 0\right|+\Omega_{21}\left|2\right\rangle_{Q Q}\hspace{-0.2em}\left\langle 1\right| \nonumber\\
& +\Omega_{31}\left|3\right\rangle_{Q Q}\hspace{-0.2em}\left\langle 1\right|+\Omega_{32}e^{i\phi_{Q}}\left|3\right\rangle_{Q Q}\hspace{-0.2em}\left\langle 2\right|+h.c.]. \label{Hamiltonian}
\end{align}
Here, $\Delta=\omega_{1}-\omega_{0}-\nu_{10}$ denotes the detuning of probe field. $\Omega_{jk}=\mu_{jk}E_{jk}/2\hbar$ describes the Rabi frequencies corresponding to the optical field $E_{jk}$ with $\mu_{jk}$ the electric-dipole transition moment of the transition $\left|j\right\rangle_{Q}$$\leftrightarrow$$\left|k\right\rangle_{Q}$. Without loss of generality, all these Rabi frequencies ($\Omega_{jk}$) have been assumed to be positive. $\phi_{Q}$ is the overall phases of the three Rabi frequencies corresponding to the cyclic three-level subsystem for the pure $R$ ($Q=R$) and $S$ ($Q=S$) enantiomers. The chirality of this model is specified by choosing the overall phases of the $R$ and $S$ enantiomers as
\begin{equation}
\phi_{R}=\phi,~\phi_{S}=\phi+\pi.\label{overphase}
\end{equation}
The dynamical evolution of the system is described by the master equations~\cite{MasterEq,Pelzer-decoherence1,Pelzer-decoherence2} as
\begin{equation}
\frac{d\rho^{Q}}{dt}=-\frac{i}{\hbar}[H_{I}^{Q},\rho^{Q}]+D(\rho^{Q}).\label{Master}
\end{equation}
Here, $D(\rho^{Q})$ denotes the decoherence~\cite{Pelzer-decoherence1,Pelzer-decoherence2} with
\begin{align}
[D(\rho^{Q})]_{jk} &=-(\gamma_{jk}+\gamma_{jk}^{\text{dph}})\rho_{jk}^{Q},~(k\neq j), \nonumber \\
[D(\rho^{Q})]_{jj} &=\underset{k^{\prime}\,(k^{\prime}>j)}{\sum}\Gamma_{k^{\prime}j}\rho_{k^{\prime}k^{\prime}}^{Q}
-\underset{k\,(k<j)}{\sum}\Gamma_{jk}\rho_{jj}^{Q}, \label{decoherenceMatrix}
\end{align}
where
\begin{equation}
\gamma_{jk}=\frac{\Gamma_{j}+\Gamma_{k}}{2},\,\Gamma_{k}=\underset{j^{\prime}\,(j^{\prime}< k)}{\sum}\Gamma_{kj^{\prime}}, \label{decoherence}
\end{equation}
where $\Gamma_{jk}$ denotes the pure population relaxation rate from state $\left|j\right\rangle $ to $\left|k\right\rangle$ resulting from spontaneous emission and inelastic collision process while $\gamma_{jk}^{\text{dph}}$ represents the pure dephasing rate arising from elastic collision process~\cite{Pelzer-decoherence1,Pelzer-decoherence2,Doyle-2012decoherenceValue}.

\subsection{Light deflection in a medium of pure enantiomers}
Light deflection phenomenon in a medium is the consequence of spatial variation of refractive index~\cite{ZhouLan-2007HomoDeflection,ZhouLan-2008HomoDeflection,Kumar-2016HomoDeflection}. In order to estimate the deflection angle of the probe light, we begin with the (general) geometrical optics differential equation in the vector form~\cite{OptPrinciples}
\begin{equation}
\frac{d}{ds}[n(\vec{r})\frac{d\vec{r}}{ds}]=\nabla n(\vec{r}),\label{trajectoryEq}
\end{equation}
where $s$ represents the length of the probe light trajectory in medium. $ds=\sqrt{dx^{2}+dz^{2}}$ is the length variation of the light ray during propagation. $\vec{r}(x,z)=x\vec{e}_{x}+z\vec{e}_{z}$ denotes a point on light ray where $\vec{e}_{x}$ and $\vec{e}_{z}$ are unit vectors along the axes. $n$ is the spatially dependent refractive index corresponding to the probe light. For simplicity, we consider that the probe light injects the medium along $z$ direction at position $\vec{r}_{0}=(x_{0},0)$. Here, we assume that the refractive index is only inhomogeneous in $x$ direction, then its gradient [i.e. $\nabla n(\vec{r})$] can reduce to $\nabla \hspace{-0.2em}_{x} n(x)$. In the small deflection angle regime, the gradient of refractive index along the probe light trajectory is approximatively equal to that at the incident position
\begin{equation}
\nabla \hspace{-0.2em}_{x} n(x) \simeq\nabla \hspace{-0.2em}_{x} n(x_{0}) \equiv n^{\prime}\vec{e}_{x},   \label{gradientApproxima}
\end{equation}
where $n^{\prime}\equiv\left.d n / d x\right|_{x=x_{0}}$ denotes the derivative of refractive index with respect to $x$. The Eq.~(\ref{trajectoryEq}) implies that the propagation trajectory of the probe light is determined by the refractive index of the medium. Thus, we then use the master equations (\ref{Master}) to derive the steady-state optical response which determines the refractive index corresponding to the probe light in a medium of pure enantiomers.

Here, we follow the method in Ref.~\cite{Perturbation} to obtain the linear optical response which can sufficiently reflect the main physical properties of the propagation of the probe light. Moreover, we now focus on the ideal zero-temperature case ($T=0\,\text{K}$), where only the ground state $\left|0\right\rangle_{Q}$ is occupied initially. In the weak probe field approximation ($\Omega_{10}\ll\Omega_{21},\Omega_{31},\Omega_{32},\Gamma_{jk},\gamma_{jk}^{\text{dph}}$) and steady-state condition (i.e., $\dot{\rho}_{jk}^{Q}=0$), the zeroth-order steady-state solution of the element $\rho_{10}^{Q}$ is zero and the first-order steady-state solution of the element $\rho_{10}^{Q}$ which determines the linear optical response for the probe light is given by (see the Appendix A)
\begin{equation}
\rho_{10}^{Q(1)}=\frac{\Omega_{10}}{\widetilde{\Omega}\text{cos}\phi_{Q}+K} \label{ProbeElement}
\end{equation}
with
\begin{align}
\widetilde{\Omega} &=\frac{2\Omega_{21}\Omega_{32}\Omega_{31}}{\kappa_{20}\kappa_{30}+\Omega_{32}^{2}}, \nonumber \\
K &=-i\frac{\kappa_{10}\Omega_{32}^{2}+\kappa_{20}\Omega_{31}^{2}+\kappa_{30}\Omega_{21}^{2}
+\kappa_{10}\kappa_{20}\kappa_{30}}{\kappa_{20}\kappa_{30}+\Omega_{32}^{2}}, \label{ElementDetail}
\end{align}
where $\kappa_{j0}=i\Delta+\gamma_{j0}+\gamma_{j0}^{\text{dph}}$~$(j=1,2,3)$. Further, the linear susceptibility of a homogeneous medium of pure enantiomers with molecular density $N_{Q}$ is defined as~\cite{QuantumOptics,
Susceptibility}
\begin{equation}
\chi_{10}^{Q}\equiv\frac{N_{Q}\left|\mu_{10}\right|^{2}}{2\hbar\varepsilon_{0}}
\cdot\frac{\rho_{10}^{Q(1)}}{\Omega_{10}},\label{linear-Susceptibility}
\end{equation}
where $\varepsilon_{0}$ represents the permittivity of vacuum. Since $|\chi_{10}^{Q}|\ll1$ in the gaseous medium whose density is low, the corresponding refractive index of the probe light is approximately determined by
\begin{equation}
n_{Q}\equiv\sqrt{1+\text{Re(\ensuremath{\chi_{10}^{Q}})}}\simeq1
+\frac{1}{2}\text{Re(\ensuremath{\chi_{10}^{Q}})}.\label{Refraction}
\end{equation}
Once the medium is driven by inhomogeneous optical fields, a spatial variation of refractive index which is essential in light deflection can be induced for the probe light. Here, we assume that $\Omega_{32}$ is homogeneous in space while $\Omega_{21}$ and $\Omega_{31}$ have the Gaussian profiles as
\begin{align}
\Omega_{21}(x)=\Omega_{21}^{(0)}\textrm{exp}(-\frac{x^{2}}{w^{2}}), \nonumber\\
\Omega_{31}(x)=\Omega_{31}^{(0)}\textrm{exp}(-\frac{x^{2}}{w^{2}}), \label{spatial-field}
\end{align}
where $\Omega_{21}^{(0)}$ [$\Omega_{31}^{(0)}$] characters the peak value of the Rabi frequencie and $w$ represents the beam radius.

Therefore, by replacing $n$ ($n^{\prime}$) in the Eqs.~(\ref{trajectoryEq}) and (\ref{gradientApproxima}) with $n_{Q}$ ($n_{Q}^{\prime}$), we can derive the propagation trajectory of the probe light which travels in the medium of pure $R$ or $S$ enantiomers~\cite{ZhouLan-2007HomoDeflection}
\begin{align}
x_{Q}(s)= &x_{0}+\frac{\ln\cosh[s\cdot n_{Q}^{\prime}]}{n_{Q}^{\prime}}, \nonumber\\
z_{Q}(s)= &\frac{\ln\sinh[s\cdot n_{Q}^{\prime}]}{n_{Q}^{\prime}}. \label{trajectory}
\end{align}
In the small deflection angle regime, the deflection angle of the probe light when it exits from the medium can be finally estimated as \cite{ZhouLan-2007HomoDeflection}
\begin{equation}
\theta_{Q}\equiv
\left.\frac{\partial_{s}x_{Q}(s)}{\partial_{s}z_{Q}(s)}\right|_{z=l_{z}}\simeq l_{z} n_{Q}^{\prime}, \label{deflectionAngle}
\end{equation}
where $l_{z}$ is the length of the medium along $z$ direction. As a result, the propagation trajectory and deflection angle of the probe light can be determined by Eqs.~(\ref{trajectory}) and (\ref{deflectionAngle}).

\subsection{Example of 1,2-propanediol}
So far, the deflection angle of the probe light in a homogeneous medium of pure enantiomers has been analyzed based on the related steady-state optical response. Specifically, in the following simulations, 1,2-propanediol~\cite{PropanediolParameter09,PropanediolParameter17} is taken as an example to demonstrate our method. In view of the spatial degeneracy of states~\cite{DegeneracyLehmann,YeChong-RealLoop,Leibscher-2019thermalpop,DegeneracyLehmannAR1,DegeneracyLehmannAR2,anglemomentum}, we choose the corresponding working states of the four-level chiral-molecule model as $|0\rangle_{Q}=|g\rangle\left|0_{0,0,0}\right\rangle$, $|1\rangle_{Q}=|e\rangle\left|1_{1,1,1}\right\rangle$, $|2\rangle_{Q}=|e\rangle\left|2_{2,1,2}\right\rangle$, and $|3\rangle_{Q}=|e\rangle\left|2_{2,0,1}\right\rangle$ where $|g\rangle$ ($|e\rangle$) denotes the corresponding vibrational ground (first-excited) state for the motion dominated by a rocking in the carbon backbone~\cite{PropanediolParameter17}. Here, the rotational states are marked in the $\left|J_{K_{a},K_{c},M}\right\rangle$ notation~\cite{DegeneracyLehmann,Leibscher-2019thermalpop,anglemomentum}. $J$ is the angular moment quantum number. $K_{a}$ runs from $J$ to $0$ and $K_{c}$ runs from $0$ to $J$ with decreasing energy~\cite{anglemomentum}. $M$ denotes the magnetic quantum number. Thus, the transitions $\left|3\right\rangle_{Q} $$\leftrightarrow$$\left|1\right\rangle_{Q}$, $\left|3\right\rangle_{Q} $$\leftrightarrow$$\left|2\right\rangle_{Q}$, and $\left|2\right\rangle_{Q} $$\leftrightarrow$$\left|1\right\rangle_{Q}$ are respectively coupled to a linearly $Z$-polarized ($\sigma=0$) and two circularly-polarized ($\sigma=\pm 1$) microwave fields whose Rabi frequencies are typical in the current experimental conditions~\cite{Doyle-2013DiscriminationLoop,Doyle-2013ChiralTWMLoop} $\left\{\Omega_{21}^{(0)}/2\pi,\Omega_{31}^{(0)}/2\pi,\Omega_{32}/2\pi\right\}\lesssim10\,\text{MHz}$. Meanwhile, the infrared transition $\left|1\right\rangle_{Q} $$\leftrightarrow$$\left|0\right\rangle_{Q}$ is coupled to a circularly-polarized ($\sigma=+1$) weak probe field. Note that the rotational transition dipole moment (of 1,2-propanediol) can be of the order of $1\,\text{D}$~\cite{PropanediolParameter09,PropanediolParameter17}, thus we here take the estimated vibrational transition dipole moment $|\mu_{10}|\sim0.1\,\text{D}$ since the dipole moment of vibrational transition is generally smaller than that of rotational transition~\cite{Leibscher-2019thermalpop}. Moreover, the bare transition frequencies are given by $\omega_{21}/2\pi=29.2087\,\text{GHz}$, $\omega_{32}/2\pi=100.76\,\text{MHz}$, and $\omega_{31}/2\pi=29.3095\,\text{GHz}$ based on the rotational constants for the vibrational first-excited state of lowest energy conformer of 1,2-propanediol  $A/2\pi=8524.405\,\text{MHz}$, $B/2\pi=3635.492\,\text{MHz}$, and $C/2\pi=2788.699\,\text{MHz}$~\cite{PropanediolParameter09,PropanediolParameter17,anglemomentum}. Meanwhile, the transition frequency $\omega_{10}/2\pi=4.328\,\text{THz}$ can be calculated based on the corresponding transition frequency between vibrational ground and first-excited states~\cite{PropanediolParameter17,anglemomentum}.

In the simulations, the decoherence, which is described by the master equation (\ref{Master}), includes these two types: pure population relaxation and pure dephasing~\cite{Doyle-2012decoherenceValue,Smith-1999decay,Shapiro-2006Decay,Pelzer-decoherence1,Pelzer-decoherence2}. Specifically, we here choose $\Gamma_{21}/2\pi=\Gamma_{32}/2\pi=\Gamma_{31}/2\pi\simeq0.06\,\text{MHz}$, $\Gamma_{10}/2\pi=\Gamma_{20}/2\pi=\Gamma_{30}/2\pi\simeq0.6\,\text{kHz}$, and $\gamma_{jk}^{\text{dph}}/2\pi\simeq0.06\,\text{MHz}~(j,k=0,1,2,3)$~\cite{vib-ro-book1,vib-ro-book2,Doyle-2012decoherenceValue,Doyle-2017decoherenceValue} according to the typical experimental conditions~\cite{Doyle-2012decoherenceValue,Doyle-2017decoherenceValue}. Moreover, the length of the medium is reasonably taken as $l_{z}=20\,\text{cm}$~\cite{Doyle-2017decoherenceValue}.

\subsection{Detecting the chirality of pure enantiomers via light deflection}
Before illustrating how to detect the chirality of pure enantiomers via monitoring the deflection angle of the probe light, we explain the physical mechanism underlying our method as following. According to Eq.~(\ref{overphase}), the media which composes of pure $R$ or $S$ enantiomers could result in chirality-dependent refractive index for the probe light [see Eq.~(\ref{Refraction})]. When two of the three strong driving fields are spatially inhomogeneous [e.g. $\Omega_{21}(x)$ and $\Omega_{31}(x)$ given in Eq.~(\ref{spatial-field})], the resultant gradient of refractive index will become chirality-dependent. In terms of the Fermat's principle~\cite{OptPrinciples}, the probe light traveling in the medium of pure enantiomers is always along the trajectory taking the least time. Thus, such a chirality-dependent gradient of refractive index can lead to completely different propagation trajectories of the probe light. As a consequence, the chirality of pure enantiomers can be distinguished by monitoring the corresponding deflection angle of the probe light in principle.

\begin{figure}[tbp]
	\includegraphics[width=8.5cm]{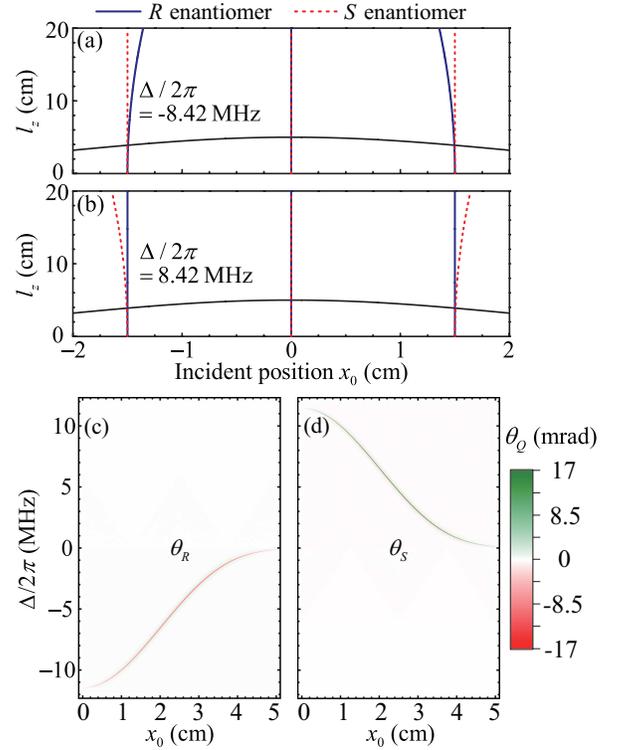}
	\caption{(Color online) The propagation trajectory of the weak probe light with detuning (a) $\Delta/2\pi=-8.42\,\text{MHz}$ and (b) $\Delta/2\pi=8.42\,\text{MHz}$ when it enters the medium at three positions: $x_{0}=-1.5\,\text{cm}$, $x_{0}=0\,\text{cm}$, and $x_{0}=1.5\,\text{cm}$. The deflection angle of the probe light corresponding to pure (c) $R$ and (d) $S$ enantiomers varies with incident position $x_{0}$ and detuning of the probe light $\Delta$. The other parameters are chosen as $\Omega_{21}^{(0)}/2\pi=\Omega_{31}^{(0)}/2\pi=10\,\text{MHz}$, $\Omega_{32}/2\pi=6\,\text{MHz}$, $|\mu_{10}|=0.1\,\text{D}$, $\phi=0$, $w=3\,\text{cm}$, $l_{z}=20\,\text{cm}$, and $N_{Q}=4\times10^{11}\,\text{cm}^{-3}$.}
	\label{fig2}
\end{figure}

In Fig.~\ref{fig2} we give the propagation trajectory of the probe light in the medium of pure $R$ or $S$ enantiomers at different incident positions $x_{0}$. Fig.~\ref{fig2}(a) shows that a blue-detuned ($\Delta/2\pi=-8.42\,\text{MHz}$) probe light experiences a ``attractive potential'' away from the central position of Gaussian driving field for the medium of pure $R$ enantiomers, while it travels almost along $z$ direction for the medium of enantio-pure $S$ enantiomers. On the contrary, it is shown in Fig.~\ref{fig2}(b) that a red-detuned ($\Delta/2\pi=8.42\,\text{MHz}$) probe light feels an ``repulsive potential'' for the medium of pure $S$ enantiomers but propagates almost along $z$ direction for the medium of pure $R$ enantiomers. This means that the two opposite enantiomers can be identified by measuring the deflection angle of the probe light. However, it is worth noting that the deflection angle is sensitive to the incident position $x_{0}$. For example, when the incident position is set to be $x_{0}=0\,\text{cm}$, neither the blue-detuned ($\Delta/2\pi=-8.42\,\text{MHz}$) {[}Fig.~\ref{fig2}(a){]} nor red-detuned ($\Delta/2\pi=8.42\,\text{MHz}$) {[}Fig.~\ref{fig2}(b){]} probe light has significant deflection when it propagates in the medium of pure $R$ or $S$ enantiomers.

To find the working region where the probe light has a significant deflection, we further display the deflection angle corresponding to pure $R$ and $S$ enantiomers versus the incident position $x_{0}$ and detuning $\Delta$ of the probe field in Fig.~\ref{fig2}(c) and Fig.~\ref{fig2}(d), respectively. Here, we only present the results in the region $x_{0}>0$ (e.g. $x_{0}=0\sim5\,\text{cm}$) since $\Omega_{21}$ ($\Omega_{31}$) is symmetric in space {[}see Eq.~(\ref{spatial-field}){]}. As one can see, if the incident position $x_{0}$ is set properly, the medium of pure $R$ enantiomers could lead to negative deflection angle at the detuning regions $\Delta/2\pi=-12 \sim 0\,\text{MHz}$ {[}Fig.~\ref{fig2}(c){]}, while the medium of pure $S$ enantiomers could result in positive deflection angle at the detuning regions $\Delta/2\pi=0 \sim 12\,\text{MHz}$ {[}Fig.~\ref{fig2}(d){]}. That means the working regions for the two opposite enantiomers are well separated. This chirality-dependent phenomenon is useful in detecting the chirality of pure enantiomers. Consequently, we should adjust the parameters and find the working regions, then the chirality of pure enantiomers can be distinguished by monitoring the corresponding sign of the deflection angle of the probe light. Additionally, one can find that the deflection angle will be larger if the probe light enters the medium at the position $x_{0}=2 \sim 3\,\text{cm}$. Hence, for simplicity, we assume the incident position of the probe light to be $x_{0}=2\,\text{cm}$ in the following discussions.

\begin{figure}[tbp]
\centering
\includegraphics[width=8.5cm]{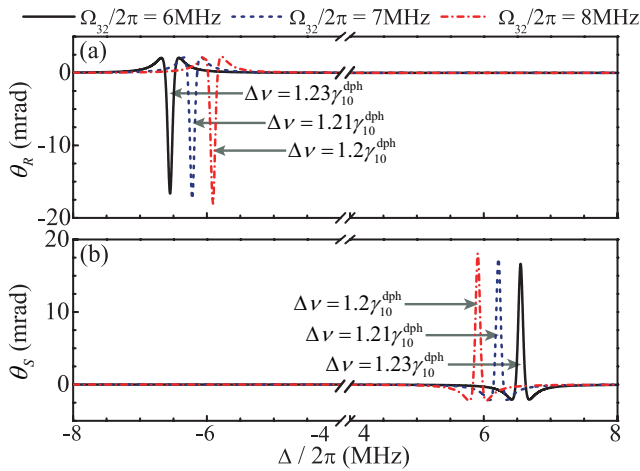}
\caption{(Color online) The deflection angle of the probe light corresponding to pure (a) $R$ and (b) $S$ enantiomers versus detuning $\Delta$ for different $\Omega_{32}$: $\Omega_{32}/2\pi=6\,\text{MHz}$, $7\,\text{MHz}$, and $8\,\text{MHz}$. Other parameters are the same as that in Fig.~\ref{fig2} except for $x_{0}=2\,\text{cm}$.}
\label{fig3}
\end{figure}

\begin{figure}[tbp]
\centering
\includegraphics[width=8.5cm]{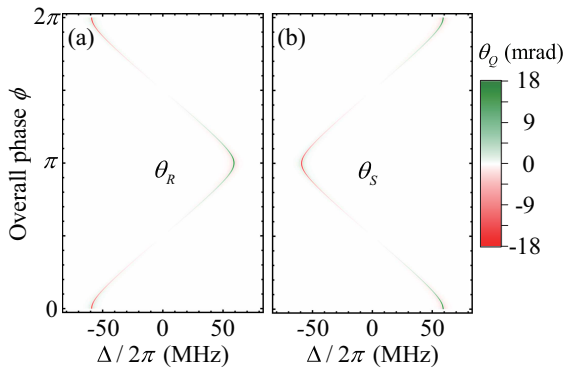}
\caption{(Color online) The deflection angle of the probe light corresponding to pure (a) $R$ and (b) $S$ enantiomers varies with overall phase $\phi$ and detuning of the probe light $\Delta$. Other parameters are the same as that in Fig.~\ref{fig3} except for $\Omega_{32}/2\pi=8\,\text{MHz}$.}
\label{fig4}
\end{figure}

The deflection angle of the probe light should be large enough to ensure that our method can act as a tool for enantio-discrimination. This immediately raises the interesting question weather the deflection angle can be controlled effectively. In Fig.~\ref{fig3}, we plot the deflection angle of the probe light corresponding to pure $R$ and $S$ enantiomers as a function of detuning $\Delta$ for different Rabi frequencies corresponding to the homogeneous driving field [$\Omega_{32}$]. One can find that the deflection angle strongly depends on $\Omega_{32}$: it increases with the enhancement of $\Omega_{32}$ due to the increase of gradient of refractive index. Namely, we can manipulate the deflection phenomenon by tuning the Rabi frequency $\Omega_{32}$ to induce larger deflection angle as required. However, $\Omega_{32}$ should be limited in the typical region of parameters [$\Omega_{32}/2\pi\lesssim10\,\text{MHz}$] according to the current experimental conditions~\cite{Doyle-2013DiscriminationLoop,Doyle-2013ChiralTWMLoop}. In recent molecular cooling experiments~\cite{Doyle-2013DiscriminationLoop,Doyle-2013ChiralTWMLoop,Doyle-2012decoherenceValue,Dopplerwidth}, molecules can be cooled to the temperature of the order of 1\,K (in the absence of microwave field). With the further development of molecular cooling technology, it is expected to prepare molecules with lower temperature. Thus, even under the heating of microwave fields, it would be still attainable to prepare molecules with temperature of the order of 1\,K. Moreover, we also give the peak width of the deflection angle $\Delta \nu$ (related to the pure dephasing rate $\gamma_{10}^{\text{dph}}$) in Fig.~\ref{fig3}. One can find that $\Delta \nu$ depends on $\Omega_{32}$: it decreases with the enhancement of $\Omega_{32}$. Note that the deflection angle here is dependent on the the first derivative of refractive index with respect to $x$ [see Eq.~(\ref{deflectionAngle})]. Thus, the lineshape of the deflection peak here is different from that of the absorption peak (whose lineshape is Lorentzian~\cite{JiaWZ-2011DiscriminationLoop}).

In realistic case, the overall phase of three microwave fields $\phi$, which has been taken to be $\phi=0$ in Figs.~\ref{fig2} and \ref{fig3}, is a tunable parameter. To investigate the effect of $\phi$ on the deflection angle, we display the deflection angle corresponding to pure $R$ and $S$ enantiomers versus the overall phase $\phi$ and detuning $\Delta$ of the probe field in Fig.~\ref{fig4}(a) and Fig.~\ref{fig4}(b). One can find that the deflection angle corresponding to pure $R$ ($S$) enantiomers, $\theta_{R}$ ($\theta_{S}$), changes from negative (positive) deflection angle to positive (negative) one gradually when $\phi$ ranges from $0$ to $\pi$. Further, when $\phi$ ranges from $\pi$ to $2\pi$, the corresponding $\theta_{R}$ ($\theta_{S}$) turns from positive (negative) deflection angle to negative (positive) one. This indicates that the chirality of pure enantiomers can also be distinguished by monitoring the deflection angle as function of detuning $\Delta$ with overall phase $\phi$ cycling, which is experimentally practical. Meanwhile, Fig.~\ref{fig4} also shows that the deflection angle will be optimal when the overall phase is taken to be $\phi=0$, $\pi$, and $2\pi$. Moreover, our scheme requires that the three microwave fields are phase locked to each other. Note that many experimental schemes have utilized such phase-locking among three microwave fields to achieve state-specific enantiomeric transfer~\cite{Doyle-2017decoherenceValue,Schnell-2017TransferLoop} based on the cyclic three-level model. Namely, the phase-locking is in fact possible given today's technology in experiments.

\section{DETECTING ENANTIOMERIC EXCESS OF CHIRAL MIXTURE}\label{detectEE}
In the above section, we have demonstrated that the chirality of pure enantiomers can be detected by monitoring the deflection angle of the probe light. Further, such a chirality-dependent deflection angle can also be utilized to detect the enantiomeric excess of chiral mixture.

\subsection{Light deflection in a medium of chiral mixture}
We now turn to consider the probe light propagating in a medium of chiral mixture. In this case, both kinds of $R$ and $S$ enantiomers in the medium have effect on the propagation trajectory of the probe light, thus the linear susceptibility of a homogeneous medium of chiral mixture with molecular density $N=N_{R}+N_{S}$ is written as
\begin{equation}
\chi_{10}=\chi_{10}^{R}+\chi_{10}^{S}, \label{Mixsuscept}
\end{equation}
where $\chi_{10}^{R}$ and $\chi_{10}^{S}$ can be derived via Eq.~(\ref{linear-Susceptibility}). Now, $N_{R}$ and $N_{S}$ are, respectively, the molecular densities of $R$ and $S$ enantiomers in the medium of chiral mixture. Based on Eq.~(\ref{Refraction}), the corresponding refractive index of the probe light is approximately equal to
\begin{equation}
n\equiv\sqrt{1+\text{Re(\ensuremath{\chi_{10}^{R}+\chi_{10}^{S}})}}\simeq 1+\frac{1}{2}
\text{Re(\ensuremath{\chi_{10}^{R}+\chi_{10}^{S}})}. \label{Mixrefraction}
\end{equation}
According to the above approximation, the derivative of refractive index with respect to $x$ is given as
\begin{equation}
n^{\prime}=n_{R}^{\prime}+n_{S}^{\prime}. \label{Mixgradient}
\end{equation}
The propagation trajectory of the probe light in the medium of chiral mixture is also described by the geometrical optics differential equation (\ref{trajectoryEq}). Therefore, based on the calculations in Sec.~\ref{detectchirality}, we can estimate the deflection angle of the probe light given by
\begin{equation}
\theta=\theta_{R}+\theta_{S} \label{ChiralMixAngle}
\end{equation}
with
\begin{equation}
\theta_{R}=l_{z}n_{R}^{\prime},\,\theta_{S}=l_{z}n_{S}^{\prime}. \label{MixAngledetail}
\end{equation}
Here, $\theta_{R}$ and $\theta_{S}$ can be used to respectively describe the contributions of $R$ and $S$ enantiomers to the deflection angle of probe light. In the next subsection, we will illustrate how to detect the enantiomeric excess of chiral mixture via such a deflection angle.

\subsection{Detecting the enantiomeric excess of chiral mixture via light deflection}
\begin{figure}[tbp]
\centering
\includegraphics[width=8.5cm]{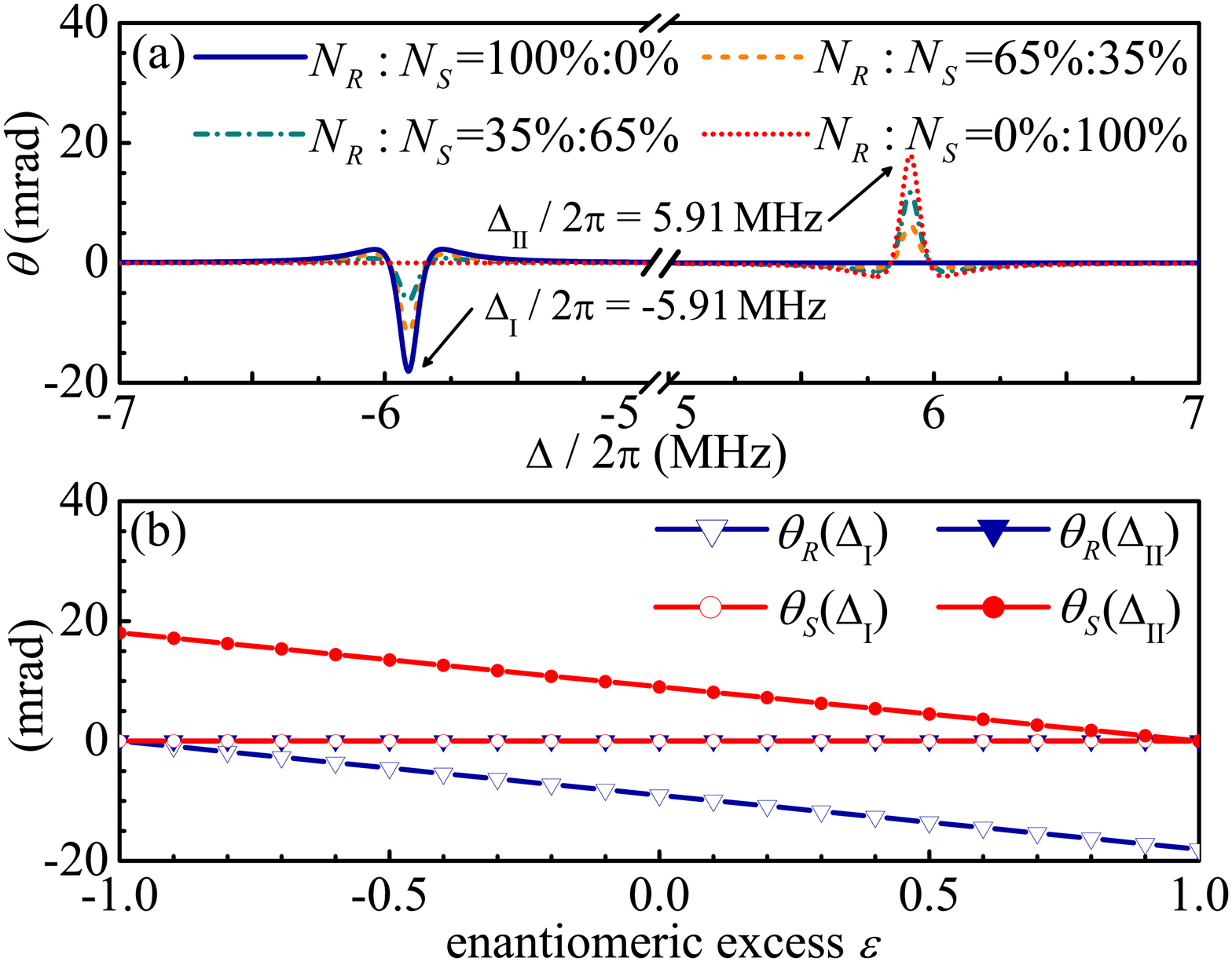}
\caption{(Color online) (a) The deflection angle of the probe light as a function of detuning $\Delta$ for different percentages of two opposite enantiomers. (b) The contributions of $R$ and $S$ enantiomers to the deflection angle of the probe light $\theta$ at two characteristic detunings $\Delta_{\text{I}}/2\pi=-5.91\,\text{MHz}$ and $\Delta_{\text{II}}/2\pi=5.91\,\text{MHz}$ as a function of the enantiomeric excess. The parameters are $N=4\times10^{11}\,\text{cm}^{-3}$, $\Omega_{32}/2\pi=8\,\text{MHz}$ and $x_{0}=2\,\text{cm}$. Other parameters are the same as that in Fig.~\ref{fig2}.}
\label{fig5}
\end{figure}

As shown in the last section, each of the two opposite enantiomers can lead to prominent light deflection at one regions {[}the region $\Delta/2\pi=-12 \sim 0\,\text{MHz}$ for pure $R$ enantiomers in Fig.~\ref{fig2}(c), and the region $\Delta/2\pi=0 \sim 12\,\text{MHz}$ for pure $S$ enantiomers in Fig.~\ref{fig2}(d){]}. In Fig.~\ref{fig5}(a), we show the deflection angle of the probe light versus the detuning of the probe light ($\Delta$) for different percentages of two opposite enantiomers in the chiral mixture. Apparently, there are two characteristic peaks of deflection angle at $\Delta/2\pi=-5.91\,\text{MHz}\equiv\Delta_{\text{I}}/2\pi$ and $\Delta/2\pi=5.91\,\text{MHz}\equiv\Delta_{\text{II}}/2\pi$. It is shown that the amplitude of deflection angle at the detuning $\Delta_{\text{I}}$ is larger (smaller) than that at $\Delta_{\text{II}}$ when the density of $R$ enantiomers ($S$ enantiomers) in the chiral mixture is dominant. These results suggest that the percentages of the two opposite enantiomers are directly associated with the deflection angles of the probe light at the two characteristic detunings. To further investigate this property, in Fig.~\ref{fig5}(b), we display $\theta_{R}(\Delta_{\text{I}})$, $\theta_{R}(\Delta_{\text{II}})$, $\theta_{S}(\Delta_{\text{I}})$, and $\theta_{S}(\Delta_{\text{II}})$ as a function of the enantiomeric excess which is defined as $\varepsilon\equiv\frac{N_{R}-N_{S}}{N_{R}+N_{S}}$. As one can see, at the detuning $\Delta_{\text{I}}$ ($\Delta_{\text{II}}$), the contributions of $R$ enantiomers ($S$ enantiomers) is prominent while that of $S$ enantiomers ($R$ enantiomers) tends to be negligible. This implies that, in appropriate region of parameters, the contributions of $R$ enantiomers ($S$ enantiomers) to the deflection angle at the detuning $\Delta_{\text{II}}$ ($\Delta_{\text{I}}$) can be approximately reduced to
\begin{equation}
\theta_{R}(\Delta_{\text{II}})\simeq0,\,\theta_{S}(\Delta_{\text{I}})\simeq0.
\label{absentGradient}
\end{equation}
Applying this approximation to Eq.~(\ref{ChiralMixAngle}), we can obtain
\begin{align}
\theta(\Delta_{\text{I}}) \simeq \theta_{R}(\Delta_{\text{I}}),\,\theta(\Delta_{\text{II}}) \simeq \theta_{S}(\Delta_{\text{II}}). \label{peakValueApproxima}
\end{align}
As a result, the enantiomeric excess of chiral mixture can be estimated as
\begin{equation}
\varepsilon\simeq\varepsilon_{m}\equiv\frac{\left|\theta(\Delta_{\text{I}})\right|
-\left|\theta(\Delta_{\text{II}})\right|}{\left|\theta(\Delta_{\text{I}})\right|
+\left|\theta(\Delta_{\text{II}})\right|},
\label{measEE}
\end{equation}
where the subscript ``$m$'' denotes the estimation value. Therefore, in order to detect the enantiomeric excess of chiral mixture, we should adjust the parameters to find two characteristic peaks for the light deflection angle at two different detunings of probe light where the corresponding amplitudes of two deflection angles are approximately proportional to the densities of $R$ enantiomers and $S$ enantiomers, respectively.

\begin{figure}[tbp]
\centering
\includegraphics[width=8.5cm]{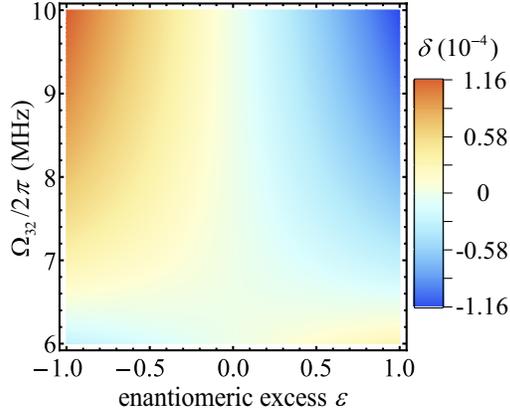}
\caption{(Color online) Robustness of the absolute error $\delta$ with respect to the Rabi frequency $\Omega_{32}$ and enantiomeric excess $\varepsilon$. Note that we should sweep the detuning of the probe light to find the new characteristic detunings $\Delta_{\text{I}}$ and $\Delta_{\text{II}}$ once $\Omega_{32}$ is changed since the variation of $\Omega_{32}$ would shift the two characteristic detunings. Other parameters are the same as that in Fig.~\ref{fig5}.}
\label{fig6}
\end{figure}

Moreover, it is worth mentioning that the approximation in Eq.~(\ref{absentGradient}) can be ensured when pure $R$ and $S$ enantiomers could respectively lead to significant deflection of the probe light at different working regions {[}see in Fig.~\ref{fig2}(c) and Fig.~\ref{fig2}(d){]}. However, if there are overlaps between the working region related to pure $R$ enantiomers and that related to $S$ enantiomers, the corresponding estimation precision will suffer from such overlap due to the failure of $\theta_{R}(\Delta_{\text{II}})\simeq0$ and $\theta_{S}(\Delta_{\text{I}})\simeq0$. Therefore, it is necessary to evaluate the estimation precision of our method. Here, we introduce the absolute error
\begin{equation}
\delta=\varepsilon_{m}-\varepsilon \label{error}
\end{equation}
to describe the estimation precision and verify the robustness of $\delta$ against a variation of the enantiomeric excess $\varepsilon$ and Rabi frequency $\Omega_{32}$ in Fig.~\ref{fig6}. One can find that an affordable absolute error (below $0.02\%$) is fairly robust against the variation of $\varepsilon$ and $\Omega_{32}$ in the range $\Omega_{32}/2\pi=6\sim10\,\text{MHz}$ and $\varepsilon=-1\sim1$.

\section{Discussion}\label{discussion}
The principle of our method is based on the light deflection effect in a homogeneous medium of chiral molecules subjected to inhomogeneous external fields. In the above simulations, the frequency of the probe field is of the order of 1\,THz. Therefore, in principle, a device sensitive to the positional change of the optical field whose frequency is of the order of 1\,THz would be required to make sure that the corresponding deflection angle can be detected.Moreover, in the above simulations, we use the geometrical optics differential equation [Eq.~(\ref{trajectoryEq})] to estimate the deflection angle of the probe field since it is a valid tool for investigate the propagation trajectory of the probe field~\cite{ZhouLan-2007HomoDeflection,Kumar-2016HomoDeflection,ZhouLan-2009DeflectionAppl,Scully-2010DeflectionAppl,ZhuChengJie-2013DeflectionAppl}. In practice, the beam size of the probe field should be an important consideration since it would lead to the broadening of the frequencies of the deflection peaks. We will consider to investigate this issue in the future work.

In the above investigation,
we only focus on the ideal zero-temperature case ($T=0\,\text{K}$), where the molecules only occupy the ground state, thus the corresponding ground-state molecular density is equal to the molecular density $N$ (similarly $N_{L}$ and $N_{R}$). In realistic experiments~\cite{Doyle-2013DiscriminationLoop,Lehmann-2015DiscriminationLoop,Schnell-2015DiscriminationLoop,Schnell-2017TransferLoop,Leibscher-2019thermalpop}, even under “cold” conditions, many molecular ro-vibrational states are thermally occupied due to the effect of environment temperature. However, note that almost only the ground vibrational state ($|g\rangle\left|J_{\tau,M}\right\rangle$) is thermally occupied in “cold” experiments~\cite{Doyle-2013DiscriminationLoop,Lehmann-2015DiscriminationLoop,Schnell-2015DiscriminationLoop,Schnell-2017TransferLoop,Leibscher-2019thermalpop} due to the large vibrational transition frequency (in the infrared range)~\cite{Leibscher-2019thermalpop}. Thus, among the four states $\left|0\right\rangle_{Q}$, $\left|1\right\rangle_{Q}$, $\left|2\right\rangle_{Q}$, and $\left|3\right\rangle_{Q}$, almost only the ground state ($\left|0\right\rangle_{Q}$) is occupied at finite temperature. Even so, under the cooled condition, there still exist other rotational states of the ground vibrational state like $|g\rangle\left|J_{\tau,M}\right\rangle\,(J\geq1)$ which are thermally occupied. For the temperature $T$, the thermal population on the rotational state of ground vibrational state with eigen-energy $\hbar\omega_{r}$~\cite{PropanediolParameter09,	PropanediolParameter17,anglemomentum} is given by $P_{r}=\exp \left(-\frac{\hbar \omega_{r}}{k_{B} T}\right) / \sum_{r} \exp \left(-\frac{\hbar \omega_{r}}{k_{B} T}\right)$ where $k_{B}$ is the Boltzmann constant, and $\sum_{r}$ denotes the sum for all rotational states (for given ground vibrational state $|g\rangle$). Specially, at typical temperature about $T=1\,\text{K}$~\cite{Doyle-2012decoherenceValue,Dopplerwidth}, the proportion of state $|g\rangle\left|0_{0,0}\right\rangle$ (i.e., the ground state of the four-level chiral-molecule model) to the total molecules due to thermal occupation is about \textcolor{blue}{$5\%$} for the ground vibrational state~\cite{PropanediolParameter09,	PropanediolParameter17}. As a result, the corresponding ground-state molecular density is less than the molecular density $N$. Based on this, to ensure that the above results (i.e., Figs.~\ref{fig2}, \ref{fig3}, \ref{fig4}, \ref{fig5}, and \ref{fig6}) can remain unchanged for typical temperature about $T=1\,\text{K}$, we should take the ground-state ($\left|0\right\rangle_{Q}=|g\rangle\left|0_{0,0}\right\rangle$) molecular density to be $4\times10^{11}\,\text{cm}^{-3}$ [i.e., the molecular density $N$ (similarly $N_{L}$ and $N_{R}$) mentioned in above discussion should be replaced by the ground-state molecular density], and thus the corresponding molecular density should be increased to about $8\times10^{12}\,\text{cm}^{-3}$ which is still attainable in today's buffer gas cooling experiment~\cite{Doyle-2013DiscriminationLoop,Doyle-2017decoherenceValue,Doyle-2012decoherenceValue}.

Moreover, in realistic case, the non-zero temperature would in fact lead to inevitable Doppler effect~\cite{Dopplerwidth,DopplerSuppress,Doppler1,Doppler2}, and then a reasonable Doppler broadening of the $\left|1\right\rangle_{Q}$$\leftrightarrow$$\left|0\right\rangle_{Q}$ transition should be taken into consideration. Spectically, for the temperature $T\simeq1\,\text{K}$~\cite{Doyle-2012decoherenceValue,Dopplerwidth}, the corresponding Doppler full width at half maximum (defined as $\Delta\nu_{D}=\omega_{10}\sqrt{\frac{8 k_{B} T \mathrm{ln}2}{M c^2}}$)~\cite{Dopplerwidth} here is about $2\pi \times0.35\,\text{MHz}$ (based on the parameters: mass of 1,2-propanediol $M=1.264 \times 10^{-25}\,\text{kg}$, central transition frequency  $\omega_{10}/2\pi=4.328\,\text{THz}$, Boltzmann constant $k_{B}$, and light speed in vacuum $c$)~\cite{PropanediolParameter09,PropanediolParameter17}. Such a Doppler broadening could normally reasult in the broadening of spectral lines which may obscure the characteristic peaks. However, by taking appropriate Rabi frequencies to ensure the characteristic peaks are separated from each other sufficiently~\cite{DopplerSuppress,JiaWZ-2011DiscriminationLoop} (i.e., ensuring $|\Delta_{\text{I}}|$~and~$|\Delta_{\text{II}}|\gg\Delta\nu_{D}$), the similar chirality-dependent light deflection could be still observed. On the other hand, the
Doppler broadening would also lead to the suppression of deflection angle since the Doppler width $\Delta\nu_{D}$ here is larger than the peak width of deflection angle $\Delta\nu \approx 2\pi \times0.07\,\text{MHz}$ [see Fig.~\ref{fig3}]. However, note that molecules can be cooled to the temperature on the order of $1\,\text{K}$ in recent molecular cooling experiments~\cite{Doyle-2013DiscriminationLoop,Doyle-2013ChiralTWMLoop,Doyle-2012decoherenceValue,Dopplerwidth}, it is expected to prepare molecules with lower temperature (below $1$\,\text{K}) in the near future, and then the corresponding effect of Doppler broadening can be suppressed further.

Finally, in the schemes for enantio-discrimination based on cyclic three-level models~\cite{JiaWZ-2011DiscriminationLoop,Hirota-2012DiscriminationLoop,Doyle-2013DiscriminationLoop,Doyle-2013ChiralTWMLoop,Doyle-2014ChiralTWMLoop,Lehmann-2015DiscriminationLoop,Schnell-2014DiscriminationLoop,Schnell-2015DiscriminationLoop,Lehmann-2017DiscriminationLoop}, phase matching among the electromagnetic fields is an important consideration~\cite{Lehmann-2017DiscriminationLoop,DegeneracyLehmann}. In practice, to ensure the overall phase $\phi$ can be approximately the same for all molecules (which means that the phase matching is fulfilled), the product of the apparatus characteristic length $l$ and phase mismatching wave vector $\Delta\vec{k}=\vec{k}_{31}-\vec{k}_{21}-\vec{k}_{32}$ should meet the requirement $|\Delta\vec{k}| l\ll2\pi$ (with $k=2\pi/\lambda$ where $\lambda$ denotes the wavelength). Following the method in Refs.~\cite{Lehmann-2017DiscriminationLoop,DegeneracyLehmann} (i.e., taking one of the transition frequencies for the three-level model as small as possible), the effect of phase mismatching here can be minimized by taking $\vec{k}_{21}$ and $\vec{k}_{31}$ to be parallel and $\vec{k}_{32}$ to be perpendicular to them. Then, for an apparatus characteristic length about $l\simeq20\,\text{cm}$ and wave vector $|\vec{k}_{32}|=2\pi \times 0.336\,\text{m}^{-1}$, the corresponding phase mismatching is $|\Delta\vec{k}| l =\sqrt{2}|\vec{k}_{32}|l\simeq0.596$ which can meet the requirement $|\Delta\vec{k}| l\ll2\pi$, thus effect of phase mismatching is negligible.

\section{CONCLUSION}\label{summary}
In conclusion, we have proposed a theoretical method for enantio-discrimination based on the light deflection effect in a four-level chiral-molecule model consisting of a cyclic three-level subsystem and an auxiliary level. The key idea is to induce spatially inhomogeneous refractive index which is chirality-dependent using inhomogeneous driving fields, and then the chirality of pure enantiomers can be mapped on the deflection angle of the probe light. We also investigate the effect of the Rabi frequency of the driving field on the deflection angle and the results show that the amplitude of deflection angle can be controlled effectively via such a Rabi frequency. Further, we demonstrate that such a chirality-dependent light deflection angle can also be utilized to detect the enantiomeric excess of chiral mixture with uniform molecular density. Moreover, we also evaluate the measurement precision of our method. It is shown that we can obtain affordable measurement precision which is fairly robust against a small variation in the Rabi frequency and enantiomeric excess. Therefore, our method may act as a tool for enantio-discrimination.

\section*{Supplementary Material}
See supplementary material for the complete calculation of the first-order steady-state solution $\rho_{10}^{Q(1)}$ which determines the linear optical response for the probe light. Moreover, the figure about the comparison between $\rho_{10}^{Q(1)}$ and $\rho_{10}^{Q}$ is given in the supplementary material.

\section*{acknowledgments}
This work was supported by the National Key R\&D Program of China (Grant No. 2016YFA0301200), the Natural Science Foundation of China (Grants No. 11774024, No. 11534002, No. U1930402, and No. 11947206), and the Science Challenge Project (Grant No. TZ2018003).

\section*{Data Availability Statements}
The data that supports the findings of this study are available within the article [and its supplementary material].

\end{document}


\title{Supplementary Material}

\maketitle

\begin{appendices}

\renewcommand{\theequation}{S\arabic{equation}}
\renewcommand{\thefigure}{S\arabic{figure}}

See supplementary material for the complete calculation about the first-order steady-state solution of the element $\rho_{10}^{Q}$ which determines the linear optical response for the probe light.
Here, by substituting the Hamiltonian (\textcolor{red}{1}) to the master equations (\textcolor{red}{3}), we can obtain the dynamical evolution equations of the system for the ideal zero-temperature case ($T=0\,\text{K}$) as
\begin{align}
	\dot{\rho}_{00}^{Q}=&i(\Omega_{10}\rho_{10}^{Q}+\Omega_{10}\rho_{01}^{Q})
	+\Gamma_{10}\rho_{11}^{Q}+\Gamma_{20}\rho_{22}^{Q}+\Gamma_{30}\rho_{33}^{Q}, \label{q00}\\
	\dot{\rho}_{11}^{Q}=&i(\Omega_{10}\rho_{01}^{Q}+\Omega_{21}\rho_{21}^{Q}+\Omega_{31}\rho_{31}^{Q}
	+\Omega_{10}\rho_{10}^{Q}+\Omega_{21}\rho_{12}^{Q}+\Omega_{31}\rho_{13}^{Q}) -\Gamma_{10}\rho_{11}^{Q}+\Gamma_{21}\rho_{22}^{Q}+\Gamma_{31}\rho_{33}^{Q}, \label{q11}\\
	\dot{\rho}_{22}^{Q}=&i(\Omega_{21}\rho_{12}^{Q}+\Omega_{32}e^{-i\phi_{Q}}\rho_{32}^{Q}
	+\Omega_{21}\rho_{21}^{Q}+\Omega_{32}e^{i\phi_{Q}}\rho_{23}^{Q})
	-\Gamma_{2}\rho_{22}^{Q}+\Gamma_{32}\rho_{33}^{Q}, \label{q22}\\
	\dot{\rho}_{33}^{Q}=&i(\Omega_{31}\rho_{13}^{Q}+\Omega_{32}e^{i\phi_{Q}}\rho_{23}^{Q}
	+\Omega_{31}\rho_{31}^{Q}+\Omega_{32}e^{-i\phi_{Q}}\rho_{32}^{Q})-\Gamma_{3}\rho_{33}^{Q}, \label{q33}\\
	\dot{\rho}_{10}^{Q}=&i(\Omega_{10}\rho_{00}^{Q}-\Omega_{10}\rho_{11}^{Q}+\Omega_{21}\rho_{20}^{Q}
	+\Omega_{31}\rho_{30}^{Q})-\kappa_{10}\rho_{10}^{Q}, \label{q10}\\
	\dot{\rho}_{01}^{Q}=&i(\Omega_{10}\rho_{11}^{Q}-\Omega_{10}\rho_{00}^{Q}-\Omega_{21}\rho_{20}^{Q}
	-\Omega_{31}\rho_{30}^{Q})-\kappa_{01}\rho_{01}^{Q}, \label{q01}\\
	\dot{\rho}_{20}^{Q}=&i(\Omega_{21}\rho_{10}^{Q}-\Omega_{10}\rho_{21}^{Q}+\Omega_{32}e^{-i\phi_{Q}}\rho_{30}^{Q})
	-\kappa_{20}\rho_{20}^{Q}, \label{q20}\\
	\dot{\rho}_{02}^{Q}=&i(\Omega_{10}\rho_{12}^{Q}-\Omega_{21}\rho_{01}^{Q}-\Omega_{32}e^{i\phi_{Q}}\rho_{03}^{Q})
	-\kappa_{02}\rho_{02}^{Q}, \label{q02}\\
	\dot{\rho}_{21}^{Q}=&i(\Omega_{21}\rho_{11}^{Q}-\Omega_{21}\rho_{22}^{Q}-\Omega_{10}\rho_{20}^{Q}
	-\Omega_{31}\rho_{23}^{Q}+\Omega_{32}e^{-i\phi_{Q}}\rho_{31}^{Q})-(\gamma_{21}+\gamma_{21}^{\text{dph}})\rho_{21}^{Q}, \label{q21}\\
	\dot{\rho}_{12}^{Q}=&i(\Omega_{21}\rho_{22}^{Q}-\Omega_{21}\rho_{11}^{Q}+\Omega_{10}\rho_{02}^{Q}
	+\Omega_{31}\rho_{32}^{Q}-\Omega_{32}e^{i\phi_{Q}}\rho_{13}^{Q})-(\gamma_{12}+\gamma_{12}^{\text{dph}})\rho_{12}^{Q}, \label{q12}\\
	\dot{\rho}_{30}^{Q}=&i(\Omega_{31}\rho_{10}^{Q}-\Omega_{10}\rho_{31}^{Q}+\Omega_{32}e^{i\phi_{Q}}\rho_{20}^{Q})
	-\kappa_{30}\rho_{30}^{Q}, \label{q30}\\
	\dot{\rho}_{03}^{Q}=&i(\Omega_{10}\rho_{13}^{Q}-\Omega_{31}\rho_{01}^{Q}-\Omega_{32}e^{-i\phi_{Q}}\rho_{02}^{Q})
	-\kappa_{03}\rho_{03}^{Q}, \label{q03}\\
	\dot{\rho}_{31}^{Q}=&i(\Omega_{31}\rho_{11}^{Q}-\Omega_{31}\rho_{33}^{Q}-\Omega_{10}\rho_{30}^{Q}
	-\Omega_{21}\rho_{32}^{Q}+\Omega_{32}e^{i\phi_{Q}}\rho_{21}^{Q})-(\gamma_{31}+\gamma_{31}^{\text{dph}})\rho_{31}^{Q}, \label{q31}\\
	\dot{\rho}_{13}^{Q}=&i(\Omega_{31}\rho_{33}^{Q}-\Omega_{31}\rho_{11}^{Q}+\Omega_{10}\rho_{03}^{Q}
	+\Omega_{21}\rho_{23}^{Q}+\Omega_{32}e^{-i\phi_{Q}}\rho_{12}^{Q})-(\gamma_{13}+\gamma_{13}^{\text{dph}})\rho_{13}^{Q}, \label{q13}\\
	\dot{\rho}_{32}^{Q}=&i(\Omega_{32}e^{i\phi_{Q}}\rho_{22}^{Q}-\Omega_{32}e^{i\phi_{Q}}\rho_{33}^{Q}-\Omega_{21}\rho_{31}^{Q}+\Omega_{31}\rho_{12}^{Q})-(\gamma_{32}+\gamma_{32}^{\text{dph}})\rho_{32}^{Q}, \label{q32}\\
	\dot{\rho}_{23}^{Q}=&i(\Omega_{32}e^{-i\phi_{Q}}\rho_{33}^{Q}-\Omega_{32}e^{-i\phi_{Q}}\rho_{22}^{Q}+\Omega_{21}\rho_{13}^{Q}+\Omega_{31}\rho_{21}^{Q})-(\gamma_{23}+\gamma_{23}^{\text{dph}})\rho_{23}^{Q}, \label{q23}
\end{align}
where $\kappa_{j0}=\kappa_{0j}^{*}=i\Delta+\gamma_{j0}+\gamma_{j0}^{\text{dph}}$~$(j=1,2,3)$. In the weak probe field approximation $\Omega_{10} \ll \left\{\Omega_{21},\Omega_{31},\Omega_{32},\Gamma_{jk},\gamma_{jk}^{\text{dph}}\right\}$ (this also implies that the density matrix element $\rho_{10}$ satisfies $|\rho_{10}|\ll1$) and steady-state condition (i.e. $\dot{\rho}_{jk}^{Q}=0$), the linear optical response for the probe light can be obtained by applying the perturbation approach to the elements $\rho_{jk}^{Q}$, which is introduced in terms of perturbation expansion~\cite{Perturbation}
\begin{equation}
\rho_{jk}^{Q}=\rho_{jk}^{Q(0)}+\rho_{jk}^{Q(1)}+\rho_{jk}^{Q(2)}+\cdots, \label{perturbation}
\end{equation}
where $\rho_{jk}^{Q(m)}~(m=0,1,...)$ is the $m$th-order solution of the element $\rho_{jk}^{Q}$. We substitute Eq.~(\ref{perturbation}) to Eqs.~(\ref{q00}-\ref{q23}) and treat the weak probe filed as a perturbation, then the first-order solutions of the elements $\rho_{jk}^{Q}$ satisfy the following equations
\begin{align}
	0=&i[\Omega_{10}\rho_{10}^{Q(0)}+\Omega_{10}\rho_{01}^{Q(0)}]+\Gamma_{10}\rho_{11}^{Q(1)}+\Gamma_{20}\rho_{22}^{Q(1)}+\Gamma_{30}\rho_{33}^{Q(1)},\label{q001}\\
	0=&i[\Omega_{10}\rho_{01}^{Q(0)}+\Omega_{21}\rho_{21}^{Q(1)}+\Omega_{31}\rho_{31}^{Q(1)}
	+\Omega_{10}\rho_{10}^{Q(0)}+\Omega_{21}\rho_{12}^{Q(1)}+\Omega_{31}\rho_{13}^{Q(1)}]
	-\Gamma_{10}\rho_{11}^{Q(1)}+\Gamma_{21}\rho_{22}^{Q(1)}+\Gamma_{31}\rho_{33}^{Q(1)},\label{q111}\\
	0=&i[\Omega_{21}\rho_{12}^{Q(1)}+\Omega_{32}e^{-i\phi_{Q}}\rho_{32}^{Q(1)}+\Omega_{21}\rho_{21}^{Q(1)}
	+\Omega_{32}e^{i\phi_{Q}}\rho_{23}^{Q(1)}]-\Gamma_{2}\rho_{22}^{Q(1)}+\Gamma_{32}\rho_{33}^{Q(1)},\label{q221}\\
	0=&i[\Omega_{31}\rho_{13}^{Q(1)}+\Omega_{32}e^{i\phi_{Q}}\rho_{23}^{Q(1)}+\Omega_{31}\rho_{31}^{Q(1)}
	+\Omega_{32}e^{-i\phi_{Q}}\rho_{32}^{Q(1)}]-\Gamma_{3}\rho_{33}^{Q(1)},\label{q331}\\
	0=&i[\Omega_{10}\rho_{00}^{Q(0)}-\Omega_{10}\rho_{11}^{Q(0)}+\Omega_{21}\rho_{20}^{Q(1)}+\Omega_{31}\rho_{30}^{Q(1)}]-\kappa_{10}\rho_{10}^{Q(1)},\label{q101}\\
	0=&i[\Omega_{10}\rho_{11}^{Q(0)}-\Omega_{10}\rho_{00}^{Q(0)}-\Omega_{21}\rho_{20}^{Q(1)}
	-\Omega_{31}\rho_{30}^{Q(1)}]-\kappa_{01}\rho_{01}^{Q(1)},\label{q011}\\
	0=&i[\Omega_{21}\rho_{10}^{Q(1)}-\Omega_{10}\rho_{21}^{Q(0)}+\Omega_{32}e^{-i\phi_{Q}}\rho_{30}^{Q(1)}]-\kappa_{20}\rho_{20}^{Q(1)},\label{q201}\\
	0=&i[\Omega_{10}\rho_{12}^{Q(0)}-\Omega_{21}\rho_{01}^{Q(1)}-\Omega_{32}e^{i\phi_{Q}}\rho_{03}^{Q(1)}]
	-\kappa_{02}\rho_{02}^{Q(1)},\label{q021}
\end{align}

\begin{align}
	0=&i[\Omega_{21}\rho_{11}^{Q(1)}-\Omega_{21}\rho_{22}^{Q(1)}-\Omega_{10}\rho_{20}^{Q(0)}-\Omega_{31}\rho_{23}^{Q(1)}
	+\Omega_{32}e^{-i\phi_{Q}}\rho_{31}^{Q(1)}]-(\gamma_{21}+\gamma_{21}^{\text{dph}})\rho_{21}^{Q(1)},\label{q211}\\	
	0=&i[\Omega_{21}\rho_{22}^{Q(1)}-\Omega_{21}\rho_{11}^{Q(1)}+\Omega_{10}\rho_{02}^{Q(0)}+\Omega_{31}\rho_{32}^{Q(1)}
	-\Omega_{32}e^{i\phi_{Q}}\rho_{13}^{Q(1)}]-(\gamma_{12}+\gamma_{12}^{\text{dph}})\rho_{12}^{Q(1)},\label{q121}\\
	0=&i[\Omega_{31}\rho_{10}^{Q(1)}-\Omega_{10}\rho_{31}^{Q(0)}+\Omega_{32}e^{i\phi_{Q}}\rho_{20}^{Q(1)}]-\kappa_{30}\rho_{30}^{Q(1)}, \label{q301}\\
	0=&i[\Omega_{10}\rho_{13}^{Q(0)}-\Omega_{31}\rho_{01}^{Q(1)}-\Omega_{32}e^{-i\phi_{Q}}\rho_{02}^{Q(1)}]
	-\kappa_{03}\rho_{03}^{Q(1)}, \label{q031}\\
	0=&i[\Omega_{31}\rho_{11}^{Q(1)}-\Omega_{31}\rho_{33}^{Q(1)}-\Omega_{10}\rho_{30}^{Q(0)}-\Omega_{21}\rho_{32}^{Q(1)}
	+\Omega_{32}e^{i\phi_{Q}}\rho_{21}^{Q(1)}]-(\gamma_{31}+\gamma_{31}^{\text{dph}})\rho_{31}^{Q(1)}, \label{q311}\\
	0=&i[\Omega_{31}\rho_{33}^{Q(1)}-\Omega_{31}\rho_{11}^{Q(1)}+\Omega_{10}\rho_{03}^{Q(0)}+\Omega_{21}\rho_{23}^{Q(1)}
	+\Omega_{32}e^{-i\phi_{Q}}\rho_{12}^{Q(1)}]-(\gamma_{13}+\gamma_{13}^{\text{dph}})\rho_{13}^{Q(1)}, \label{q131}\\
	0=&i[\Omega_{32}e^{i\phi_{Q}}\rho_{22}^{Q(1)}-\Omega_{32}e^{i\phi_{Q}}\rho_{33}^{Q(1)}-\Omega_{21}\rho_{31}^{Q(1)}
	+\Omega_{31}\rho_{12}^{Q(1)}]-(\gamma_{32}+\gamma_{32}^{\text{dph}})\rho_{32}^{Q(1)}, \label{q321}\\
	0=&i[\Omega_{32}e^{-i\phi_{Q}}\rho_{33}^{Q(1)}-\Omega_{32}e^{-i\phi_{Q}}\rho_{22}^{Q(1)}+\Omega_{21}\rho_{13}^{Q(1)}
	+\Omega_{31}\rho_{21}^{Q(1)}]-(\gamma_{23}+\gamma_{23}^{\text{dph}})\rho_{23}^{Q(1)}. \label{q231}
\end{align}
Here, we assume that the initial state is the ground one $\left|0\right\rangle_{Q}$. Thus the zeroth-order solutions of density matrix elements are given by
\begin{align}
	\rho_{00}^{Q(0)}&=1,~\rho_{11}^{Q(0)}=\rho_{22}^{Q(0)}=\rho_{33}^{Q(0)}=0, \label{pop}\\
	\rho_{jk}^{Q(0)}&=0 ~(k\neq j). \label{coherence}
\end{align}
By substituting the zeroth-order solutions (\ref{pop}) and (\ref{coherence}) into Eqs.~(\ref{q001}-\ref{q231}), the first-order steady-state solutions of the elements $\rho_{10}^{Q}$ is given as
\begin{equation}
\rho_{10}^{Q(1)}=\frac{\Omega_{10}}{\widetilde{\Omega}\text{cos}\phi_{Q}+K} \label{probe}
\end{equation}
with
\begin{equation}
\widetilde{\Omega}=\frac{2\Omega_{21}\Omega_{32}\Omega_{31}}{\kappa_{20}\kappa_{30}+\Omega_{32}^{2}},\,
K=-i\frac{\kappa_{10}\Omega_{32}^{2}+\kappa_{20}\Omega_{31}^{2}+\kappa_{30}\Omega_{21}^{2}
	+\kappa_{10}\kappa_{20}\kappa_{30}}{\kappa_{20}\kappa_{30}+\Omega_{32}^{2}}. \label{Detail}
\end{equation}
Further, the linear susceptibility of a homogeneous medium of pure enantiomers with molecular density $N_{Q}$ is defined as~\cite{QuantumOptics,Susceptibility}
\begin{equation}
\chi_{10}^{Q}\equiv\frac{N_{Q}\left|\mu_{10}\right|^{2}}{2\hbar\varepsilon_{0}}
\cdot\frac{\rho_{10}^{Q(1)}}{\Omega_{10}},\label{linear-Susceptibility}
\end{equation}
where $\varepsilon_{0}$ represents the permittivity of vacuum. Since $|\chi_{10}^{Q}|\ll1$ in the gaseous medium whose density is low, the corresponding refractive index of the probe light is approximately determined by
\begin{equation}
n_{Q}\equiv\sqrt{1+\text{Re(\ensuremath{\chi_{10}^{Q}})}}\simeq1
+\frac{1}{2}\text{Re(\ensuremath{\chi_{10}^{Q}})}.\label{Refraction}
\end{equation}
So far, the chirality-dependent refractive index has been obtained based on the master equations.

\setcounter{figure}{0} 
\begin{figure}[tbp]
	\centering
	\includegraphics[width=15cm]{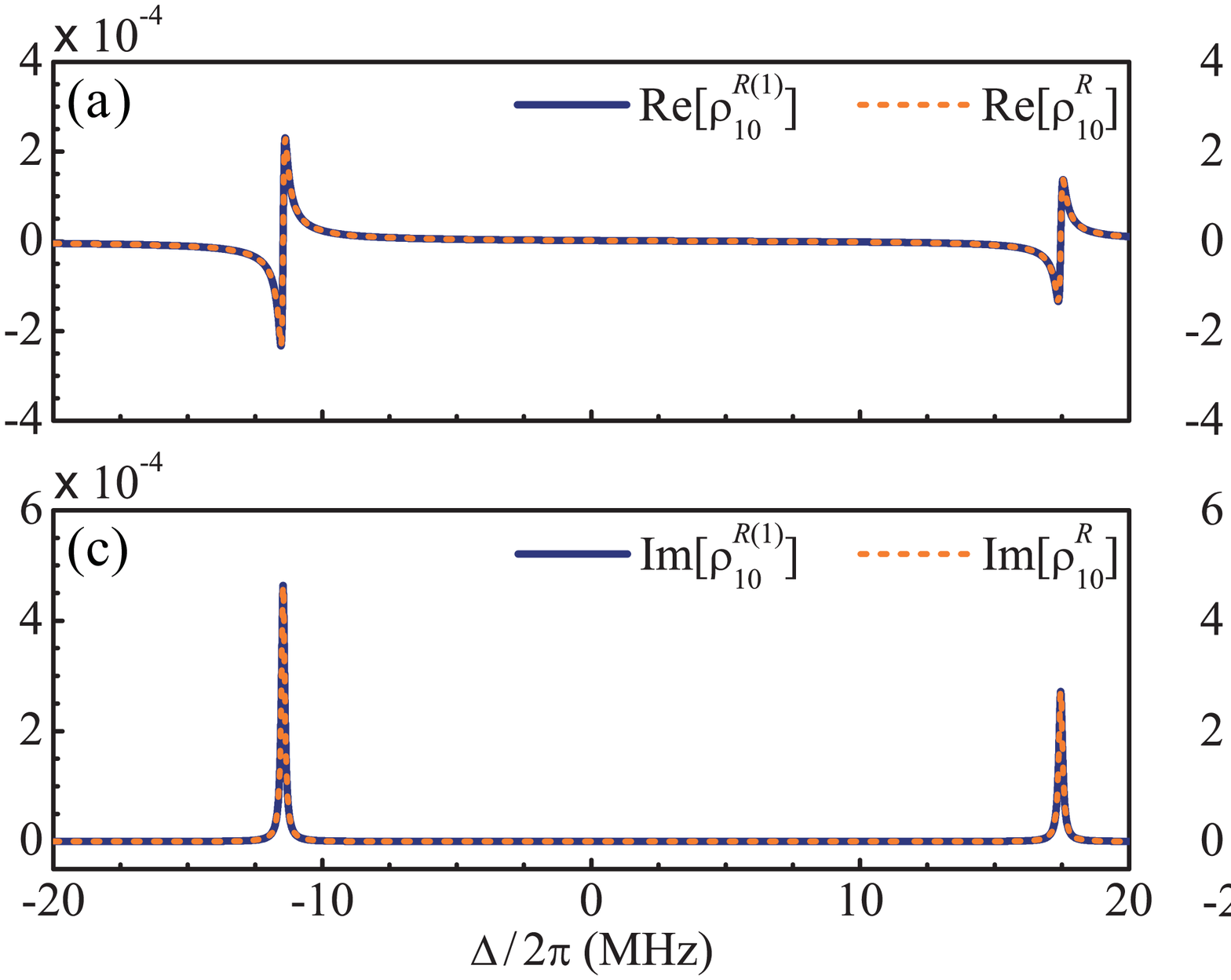}
	\caption{(Color online) (a) and (b) Real parts of $\rho_{10}^{Q}$ along with $\rho_{10}^{Q(1)}$ for $R$ and $S$ enantiomers as a function of the detuning $\Delta$. (c) and (d) Imaginary parts of $\rho_{10}^{Q}$ along with $\rho_{10}^{Q(1)}$ for $R$ and $S$ enantiomers as a function of detuning $\Delta$. Here, $\rho_{10}^{Q}$ is obtained by solving Eqs.~(\ref{q00}-\ref{q23}) numerically while $\rho_{10}^{Q(1)}$ [i.e., Eq.~(\ref{probe})] is derived from Eqs.~(\ref{q001}-\ref{coherence}). The parameters are are chosen as $\Omega_{21}^{(0)}/2\pi=\Omega_{31}^{(0)}/2\pi=10\,\text{MHz}$, $\Omega_{32}/2\pi=6\,\text{MHz}$, $\Omega_{10}/2\pi=0.06\,\text{kHz}$, $w=3\,\text{cm}$, and $x_{0}=0\,\text{cm}$.}
	\label{fig7}
\end{figure}

As shown in the above calculations, in the weak probe field approximation ($\Omega_{10} \ll \left\{\Omega_{21},\Omega_{31},\Omega_{32},\Gamma_{jk},\gamma_{jk}^{\text{dph}}\right\}$), the first-order steady-state solution of the element $\rho_{10}^{Q(1)}$ can be calculated by using the perturbation expansion (\ref{perturbation}). To illustrate the approximate analytical solution in Eq.~(\ref{probe}) $\rho_{10}^{Q(1)}$ can reflect the main property of $\rho_{10}^{Q}$, in Fig.~\ref{fig7} we show the real and imaginary parts of $\rho_{10}^{Q}$ [by numerically solving Eqs.~(\ref{q00}-\ref{q23})] and $\rho_{10}^{Q(1)}$ [according to the approximate solution in Eq.~(\ref{probe})] for $R$ and $S$ enantiomers as a function of detuning $\Delta$. One can find that the first-order steady-state solution $\rho_{10}^{Q(1)}$ is approximatively equal to $\rho_{10}^{Q}$ in the weak probe field approximation.

\end{appendices}